\documentclass[11pt]{article}
\usepackage{amsmath,amssymb,amstext,enumerate,float,graphicx}
\usepackage{hhline,lscape,syntonly,theorem,verbatim}
\usepackage{natbib}
\renewcommand{\baselinestretch}{1.0}

\newcommand{\custcap}[1]{%
\renewcommand{\baselinestretch}{1.0}
\caption{\tenrm #1}
\renewcommand{\baselinestretch}{1.0}
\normalsize}
\bibpunct{(}{)}{,}{a}{}{,}
\newcommand{\tenrm}{\fontsize{10}{12pt}\normalfont\rmfamily}
\newcommand{\upd}{\mbox{d}}

\def\exp{\mbox{e}}

\begin{document}
\vspace*{0.2in}
\bfseries

\noindent
\scalebox{1.1}{\Large The persistence of viscous effects in the
overlap region,}\\[0.06in]
\scalebox{1.1}{\Large and the mean velocity in turbulent pipe and}\\[0.06in]
\scalebox{1.1}{\Large channel flows}\\[0.12in]
\bfseries\large
Katepalli R. Sreenivasan and Anupam Sahay\\
\mdseries\large
\textit{Mason Laboratory, Yale University}\\
\textit{New Haven, CT 06520-8286}\\
\vspace*{0.08in}

\tenrm
\section*{Abstract}
We expand on our previous argument \cite[]{sreeni87} that important
elements of the dynamics of wall-bounded flows reside at the
wall-normal position $y_p$ corresponding to the peak of the Reynolds
shear stress. Specializing to pipe and channel flows, we show that
the mean momentum balance in the neighborhood of $y_p$ is distinct in
character from those in the classical inner and outer layers. We
revisit empirical data to confirm that $y_p = O((h\nu/U_*)^{1/2})$
and show that, in a neighborhood of order $R_*^{1/2}$ around $y_p$,
only the viscous effects balance pressure-gradient terms. Here, $R_*
= hU_*/\nu$, $h$ is the pipe radius or channel half-width,
$\nu$ is the kinematic viscosity of the
fluid and $U_*$ is the friction velocity. This observation provides a
mechanism by which viscous effects play an important role in regions
traditionally thought to be inviscid or inertial; in particular, it
throws doubt on the validity of the classical matching principle.
Even so, it is shown that the classical semi-logarithmic behavior for
the mean velocity distribution can be a valid approximation. It is
argued that the recently advanced power-law profiles possess a rich
underlying structure, and could be good approximations to the data
over an extended region (but they too are unlikely to be exact).

\normalsize
\section{Introduction}
Essentially all the important notions in wall-bounded flows are cast
in terms of two length scales: the inner or viscous scale $\nu/U_*$,
where $\nu$ is the kinematic viscosity of the fluid and $U_{*}$ is
the friction velocity ($\equiv\sqrt{\tau_{w}/\rho}$, $\tau_{w}$ and
$\rho$ being the wall shear stress and fluid density, respectively),
and the outer scale $h$, where $h$ is the pipe radius, channel
half-width or boundary layer thickness. The ratio of the two scales
is the Reynolds number $ R_* = hU_*/\nu$. Since the appropriate
asymptotics correspond to the limit $R_* \rightarrow \infty$, one may
expect that the problem has elements of singular perturbation. We
shall use the standard notation that $U^+ = U/U_*$ and $y^+ =
yU_*/\nu$, $y$ being the normal distance from the wall. It is
traditionally thought that the viscous effects are important up to a
$y^+$ of about 30 and that, within this region, the outer length
scale $h$ is unimportant. In the bulk of the flow excluding this
viscous region, it is thought that $\nu$ is unimportant and the
characteristic length is $h$. This view has been quite successful in
organizing various experimental data
(see, e.g., \citealt{sreeni89,dussauge96}), though it has been
recognized for some time~\cite[e.g.,][]{rao71} that the
interaction between the two scales is the key to the flow structure.

In an earlier paper \cite[]{sreeni87}, we noted that a proper
understanding of the boundary layer structure requires greater
emphasis on wall-normal position where the Reynolds shear stress
peaks. This peak position scales as the geometric mean of the inner
and outer length scales. It is well known \cite[e.g.,][]{drazin81} that,
in linear and early stages of nonlinear instability in boundary
layers and channel flows, the position of the peak Reynolds shear stress
coincides with the critical layer. This observation was inverted in
\cite{sreeni87} to suggest that the position of the peak Reynolds
shear stress, $y_p$, in turbulent wall flows plays something of the
same role as that of the critical layer in unstable wall flows. It
was pointed out that, just as for critical layer in unstable
wall flows, the mean velocity at $y_p$ is approximately a constant
fraction of the freestream or centerline velocity, the fraction being
about 0.65. Other analogies between the critical layer in the
unstable state and the `critical layer' in the turbulent state were
also cited.

The critical layer in unstable boundary layers is the seat of
perturbation vorticity which undergoes amplification when the
Reynolds number exceeds a certain threshold. The next stages of the
perturbation development involve the onset of three dimensionalities
and, eventually, of turbulence itself. If the turbulent vorticity in
the boundary layer can be caricaturized as a vortex sheet, it was
thought by analogy that its seat would be $y_p$. By interaction with image
vorticity (invoked to mimic the presence of the wall), this
hypothetical vortex sheet located at $y_p$ gets lumped first into
two-dimensional rolls and, eventually, into horse-shoe shaped
vortices. These latter structures are comparable in several respects
to those found from visualization studies of the
boundary layer~\cite[e.g.,][]{head81}. It
was further argued, albeit with less certainty, that the same picture
can explain aspects of the structure in the wall region, for example
the (noisy) spanwise periodicity of
streaks~\cite[]{kline67}. For quantitative details
and comparisons with data, one should consult \cite{sreeni87}.

Whatever the detailed objections to the physical content of the model
and however preliminary the attempt, it appeared that a
self-contained picture of the boundary layer could be developed on
that basis. Unlike in the predecessor paper \cite[]{sreeni87} which
focused on the large-scale structural elements of the boundary layer,
we shall examine here the mean velocity distribution to reiterate, in
quite a different way, the importance of the `critical layer'. This
seems to be an especially timely goal because of the renewed interest
and recent controversy surrounding the mean velocity distribution in
pipe flows \cite[e.g.,][]{barenblatt96,zagarola97}.

In the classical picture, one
{\em assumes} the existence of a common region of validity of the
outer and inner solutions of $\partial U^{+}(R_{*}, y^{+})/\partial
y^{+}$
in the limit $R_{*}\rightarrow\infty, y^{+}\rightarrow\infty$, and
puts forth asymptotic arguments to obtain
\begin{equation}
U^{+}(R_{*},y^{+}) = (1/\kappa)\ln y^{+} +B,
\end{equation}
where $\kappa$ and $B$
are empirical constants presumed to be independent of $R_{*}$. This
is the celebrated log-law, which occupies a central place in the
turbulence literature \cite[e.g.,][]{coles69,monin71,tennekes72}.
It is generally thought that
$y^{+}\approx 30$ and $y^{+} \approx 0.15 R_{*}$
are the lower and upper limits of the logarithmic profile
and the K\'{a}rm\'{a}n constant $\kappa\approx$ 0.4 and $B \approx
5.5$ (see, for example, \citealt{coles69}). There is still some
uncertainty about these constants: recent high-Reynolds-number
measurements in pipe flows~\cite[]{zagarola96} yield $\kappa = 0.44$ and
$B = 6.1$.

There exist alternative formulations for the overlap region
\cite[]{long81,barenblatt93,barenblatt96,george96a,george96b}.
Here, we shall restrict attention to Barenblatt's formulation.
Its primary contention is that the limit of
small-viscosity (or high Reynolds number)
is singular---as is common in second-order phase transitions in
condensed matter \cite[]{domb76} and also, perhaps, in Kolmogorov
turbulence \cite[]{monin75}---and so the viscous effects never disappear
in the overlap region. This imperils the classical matching argument
and the orthodox view that the log-law is exact in the
infinite Reynolds number
limit. Specifically,
note that dimensional considerations allow us to write the velocity
distribution in an intermediate layer in the form
\begin{equation}
y^+ (\partial U^{+}/\partial y^{+}) =  \psi (y^+, R_*),
\end{equation}
where $\psi$ is an unknown function of its arguments. In the
classical picture, $\psi$ is thought to asymptote to a constant, say
$1/\kappa$, as the arguments of $\psi$, namely $R_*$ and
$y^+$, assume large values. Integration then yields the log-law. On
the other hand, suppose that
\begin{equation}
y^+ (\partial U^{+}/\partial y^{+}) = (1/\kappa) (y^+)^{\alpha} ,
\end{equation}
$\alpha$ being some positive constant. This leads to a power-law for
the mean velocity distribution. In particular,
\cite{barenblatt93} and~\cite{barenblatt96} predict the asymptotic
nature of $U^{+}$ in two regions of the flow---the classical overlap
region (say B1) and a region further out towards the center (say B2).
In the power-law paradigm, B1 + B2 together form the overlap region.
The specific predictions are the following: (a) In B1, $U^{+}$ is
tangent to the classical logarithmic profile to which it remains
close but from which it always remains distinct. (b) In B2, the power
law can be approximated properly by a logarithmic function similar to
the classical log-law but with a slope that is approximately
$\sqrt{\mbox{e}}$ times that of the classical value.
It may be thought that this latter prediction does not contradict
the classical log-law because the domains of the two
logarithmic
regions are disjoint. However, because B2 would be a part of the
outer region in the classical picture, one may consider that a
conflict does exist here as well.

The purpose of this paper is to examine the nature of the mean
velocity distribution briefly, emphasizing along the way two
significant qualitative issues: (a) The viscous effects are important
in a region of pipe and channel flows that is traditionally thought
to be inviscid and that, in fact, the balance there is between
viscous and pressure gradient effects. Following~\cite{long81},
we might call this `critical' region a {\em mesolayer} (although we do not
necessarily subscribe to all the implications of that work). The
existence of such a mesolayer gives a new twist to the dynamics of
the boundary layer, but the degree to which the classical picture
needs modification is not yet clear (see section 4).  (b) The
importance of viscosity in the mesolayer offers a key to the
regeneration mechanism of the boundary layer. The discussion here
will be necessarily brief, and more details will be published
elsewhere~\cite[]{sahay96b}.

\section{The background}

\subsection{The wall-normal position of the peak of the Reynolds
shear stress}
%
\newcommand{\plfig}[3]{
\begin{tabular}{c@{\hspace{1ex}}c}
\rotatebox[origin=c]{90}{#3} &
\raisebox{-0.5\height}{\makebox{\includegraphics[trim=15 0 10 0,%
width=0.85\textwidth,height=0.36\textheight]{#1}}}\\
 & #2
\end{tabular}}

\begin{figure}[b]
\centering
\begin{picture}(420,245)(20,10)
\put(40,120){%
\mbox{\plfig{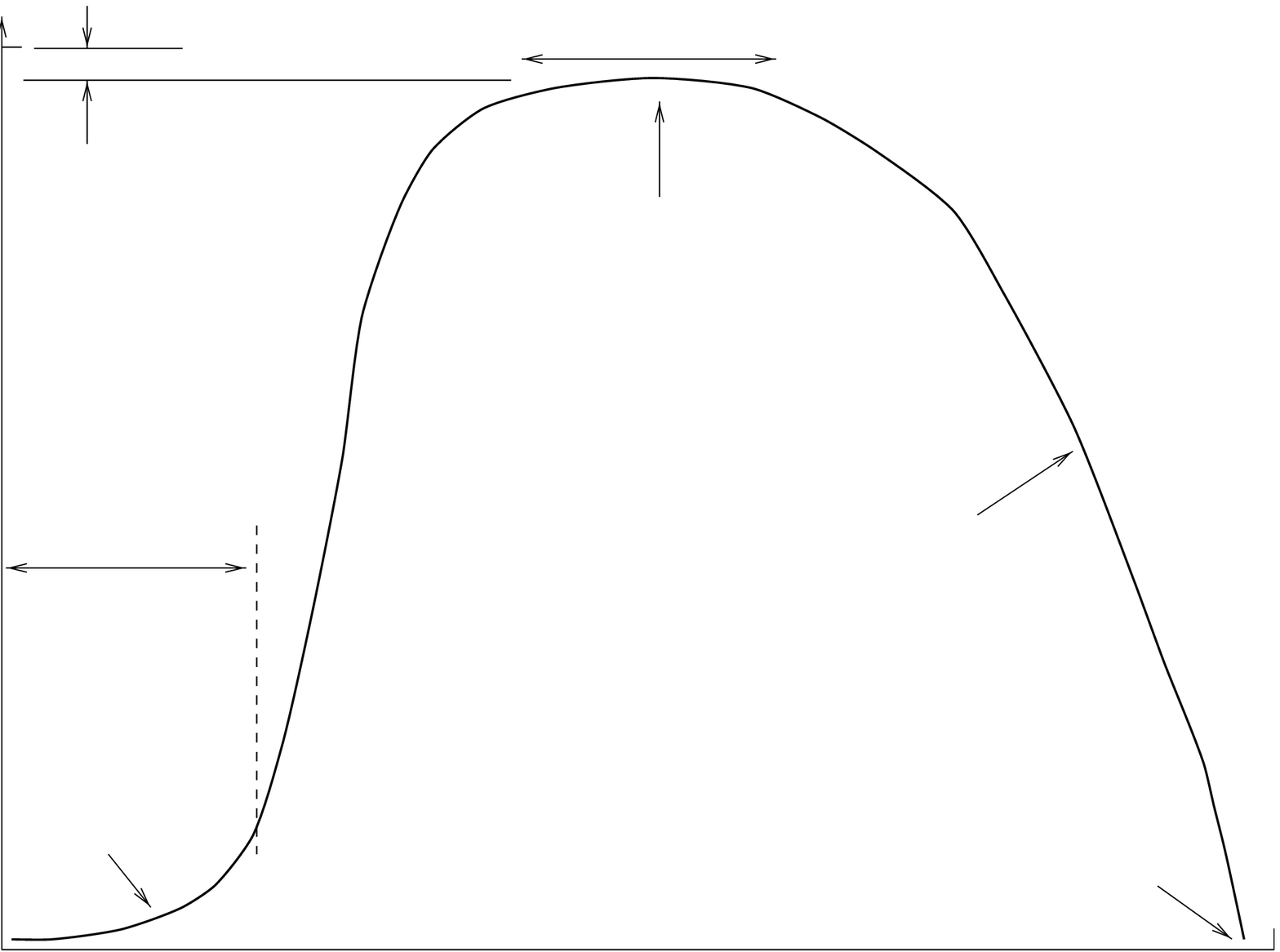}{}{}}}
\put(15,140){\scalebox{1.4}{$\tau^{+}$}}
\put(395,015){\scalebox{1.15}{$h^{+}$=$R_{*}$}}
\put(33,222){\scalebox{1.2}{1}}
\put(33,030){\scalebox{1.2}{0}}
\put(213,230){\scalebox{1.15}{slope $\approx$ 0}}
\put(213,245){\scalebox{1.15}{region (II)}}
\put(219,180){\scalebox{1.3}{$y^{+}=y_{p}^{+}$}}
\put(55,60){\scalebox{1.3}{$\sim {y^{+}}^3$}}
\put(75,200){\scalebox{1.15}{$\Delta(R_{*})$}}
\put(47,140){\scalebox{1.15}{viscous region}}
\put(70,125){\scalebox{1.15}{(I)}}
\put(265,115){\scalebox{1.15}{linear region}}
\put(285,100){\scalebox{1.15}{(III)}}
\put(335,45){\scalebox{1.15}{centerline}}
\put(200,05){\scalebox{1.3}{$\mbox{log}\;y^{+}$}}
\end{picture}
\custcap{A schematic of the turbulent shear stress $\tau^{+}$
profile in channel and pipe flows, showing three distinct regions.
The buffer layer is interposed between regions I and II.}
\label{fig:uvprof}
\end{figure}
\normalsize

Central to the present arguments is the manner in which the Reynolds
or turbulent shear stress, $\tau \equiv -\langle uv \rangle$, is
distributed in the flow; here $u$ and $v$ are velocity fluctuations
in the streamwise coordinate $x$ and the wall-normal coordinate $y$,
respectively, and $\langle \cdot \rangle$ denotes a suitable average.
Its behavior in pipe and channel flows is shown
qualitatively in Fig. 1. The quantity $\tau^+ \equiv  \tau/U_*^2$
increases from its zero value at the wall,
apparently like $y^3$ for small $y$, and rapidly reaches about half
the maximum value at a $y^+$ of about 12. It continues to increase
further to reach a maximum value at $y_p$. The maximum value equals
$\tau_w$ in the limit $R_* \rightarrow \infty$, but falls short of
$\tau_w$ at any finite $R_*$, by an amount, say, $\Delta(R_*)$. The
Reynolds stress $\tau$ decreases beyond $y_p$, and reaches zero in
the freestream of the boundary layer, and on the centerline of the
pipe or channel.

Of special importance is the position $y_p^+$.
Its leading order variation has been obtained
empirically by \cite{long81}
and \cite{sreeni87}, who have shown that
\begin{equation}
y_{p}^{+}=\lambda R_{*}^{1/2},
\end{equation}
where $\lambda$ = 1.87 and 2, respectively.  Although the two
prefactors are somewhat different, they agree on the principal result
that {\it the peak of the Reynolds shear stress occurs at a
$y^+$ that increases as $R_*^{1/2}$.} We revisit this issue here.
Figure~2a
shows plots of $\tau^+~ vs~y^{+}/R_{*}^{1/2}$ for a range of Reynolds
numbers.
The $R_*^{1/2}$ variation of $y_{p}^{+}$ appears to be a
good leading order approximation.
There is some correction to this scaling at the lowest $R_{*}$, which
we shall examine subsequently.
For the present, we have ignored the low $R_{*}$ data in estimating
$\lambda$.
On the basis of Fig.~2a, we take
$R_{*} \le 500$ as an operational definition of the low Reynolds
number. Figure~2b is an expanded plot near the peak of $\tau^+$. An
accurate determination of $y_{p}^{+}$ for large $R_{*}$ is
difficult because the peak is rather flat (and becomes more so
with increasing $R_{*}$) and because there is much scatter in the data.
Mindful of
these uncertainties we estimate that $\lambda = 1.8\pm0.2$.
%
\renewcommand{\plfig}[3]{
\begin{tabular}{c@{\hspace{1ex}}c}
\rotatebox[origin=c]{0}{#3} &
\raisebox{-0.5\height}{\makebox{\includegraphics[trim=10 0 10 0,%
width=0.9\textwidth,height=0.35\textheight]{#1}}}\\
 & #2
\end{tabular}}

\begin{figure}
\centering
\begin{picture}(420,420)(50,20)
\put(60,340){%
\mbox{\plfig{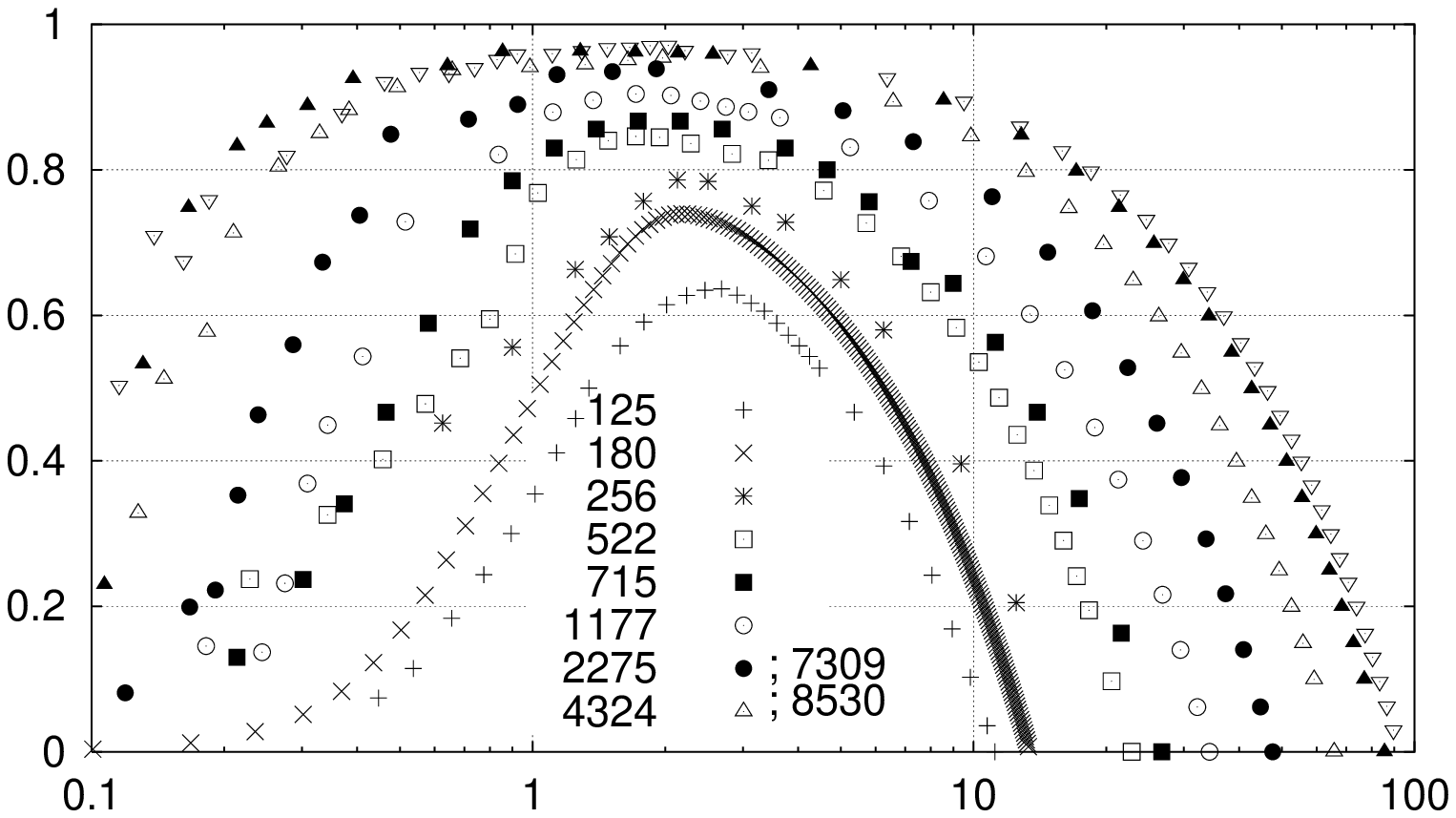}{\scalebox{1.15}{$y^{+}/R_{*}^{1/2}$}}
{}}}
\put(60,340){\scalebox{1.25}{$\tau^{+}$}}
\put(260,360){\scalebox{1.1}{$R_{*}$}}
\put(325,286){\scalebox{0.53}{$\blacktriangle$}}
\put(325,277){\scalebox{0.53}{$\triangledown$}}
\put(410,420){\scalebox{1.1}{(\textit{a})}}
\put(410,065){\scalebox{1.1}{(\textit{b})}}
\put(60,120){%
\mbox{\plfig{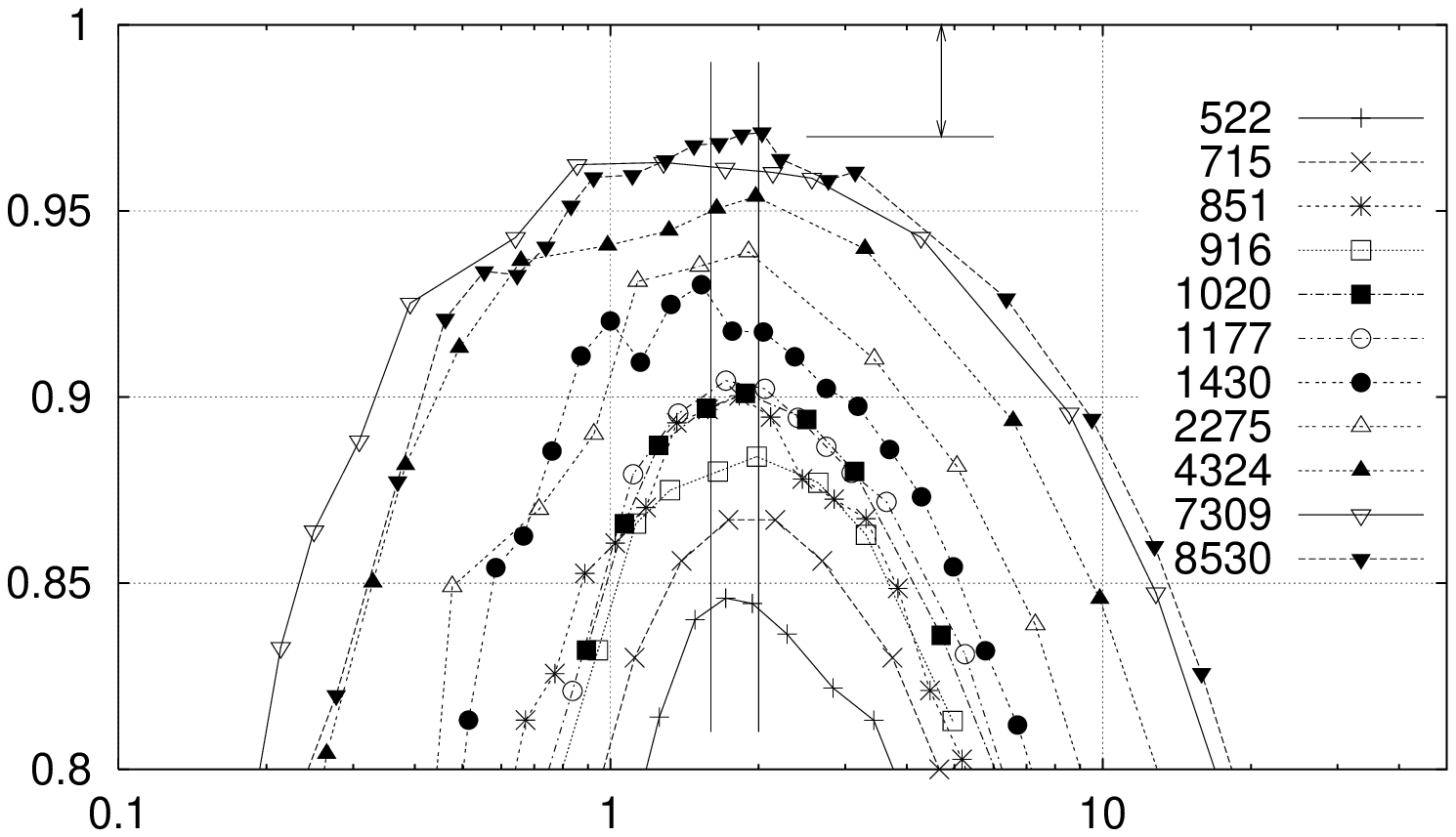}{\scalebox{1.15}{$y^{+}/R_{*}^{1/2}$}}{}}}
\put(60,120){\scalebox{1.25}{$\tau^{+}$}}
\put(294,207){\vector(-1,1){12}}
\put(258,207){\vector(1,1){12}}
\put(296,205){\scalebox{1.0}{2.0}}
\put(250,197){\scalebox{1.0}{1.6}}
\put(395,208){\scalebox{1.1}{$R_{*}$}}
\put(120,203){\scalebox{1.1}{$y_{p}^{+}=\lambda R_{*}^{1/2}$}}
\put(120,190){\scalebox{1.1}{$\lambda=1.8\pm0.2$}}
\put(335,205){\scalebox{1.0}{$\Delta(R_{*})$}}
\end{picture}
\custcap{Plots of the turbulent shear stress $\tau^{+}$ as a
function of $y^{+}/R_{*}^{1/2}$ (\textit{a})
across the channel and (\textit{b}) near its peak for high
Reynolds number experiments. The sources for
the experimental data are Antonia \textit{et al.} 1992
(channel, $R_{*}=256,\:916$), Comte-Bellot 1963 (channel, $R_{*}=4324,\:7309$),
Kim \textit{et al.} 1987 (channel DNS, $R_{*}=180$),
Laufer 1950 (channel, $R_{*}=522,\:1177,\:2275$),
Laufer 1954 (pipe, $R_{*}=8530$), Sirovich \textit{et al.} 1991
(channel DNS, $R_{*}=125$), Wei \& Willmarth 1989
(channel, $R_{*}=715,\:1020$), and Zagarola 1996 (pipe,
$R_{*}=851,\:1430$).
The shear stress has been obtained
by the numerical differentiation of the measured velocity profile
using Eq.~(\ref{eq:meanmom}) of section 3. Zagarola's data for higher Reynolds
numbers could not be used because the mean velocity data have not
been measured close enough to the wall.}
\label{fig:uvtop}
\end{figure}
%

The point to emphasize is that, for all but the very low
$R_*$, $y_p^+ = O(R_*^{1/2})$ lies well within the classical
logarithmic region ($30  \le y^+ \le 0.15 R_*$).  We shall now
discuss its role in determining the distribution of the mean velocity
in boundary layer flows. The discussion is specialized for analytical
convenience to plane channel flow and axisymmetric pipe flow. The
simplicity to be gained is that all the terms in the mean momentum
equation are independent of the streamwise direction.

\subsection{The basic physical idea}
The exact mean momentum equation is given by
\begin{equation}
-\frac{\upd P} {\upd x}+ \frac{\upd \tau}{\upd y} +
\nu \frac {\upd^2 U}{\upd y^2} = 0,
\label{eq:mombal}
\end{equation}
where $P$ is the mean pressure and U is the mean velocity depending
only on $y$.
The Reynolds shear stress term appears in the equation as an unknown.
We now make the obvious point that, at the position at which the
turbulent shear stress $\tau$ is a maximum, i.e., at $y_p$, the
pressure gradient terms are balanced only by viscous terms; the
Reynolds stress terms are entirely absent because {\it the quantity
that appears in the momentum equation is the Reynolds shear stress
gradient, not the Reynolds shear stress itself.}  We have already
seen that $y_p$ resides in the part of the boundary layer
traditionally thought to be independent of viscosity, or purely
inertial. This means that, {\it within a region in pipes and channels
that has been thought to be inertial, there exists a neighborhood
within which only viscous terms are capable of balancing the pressure
gradient terms nearly entirely.}

%
\begin{figure}[t]
\centering
\begin{picture}(420,300)(20,20)
\put(20,210){%
\mbox{\plfig{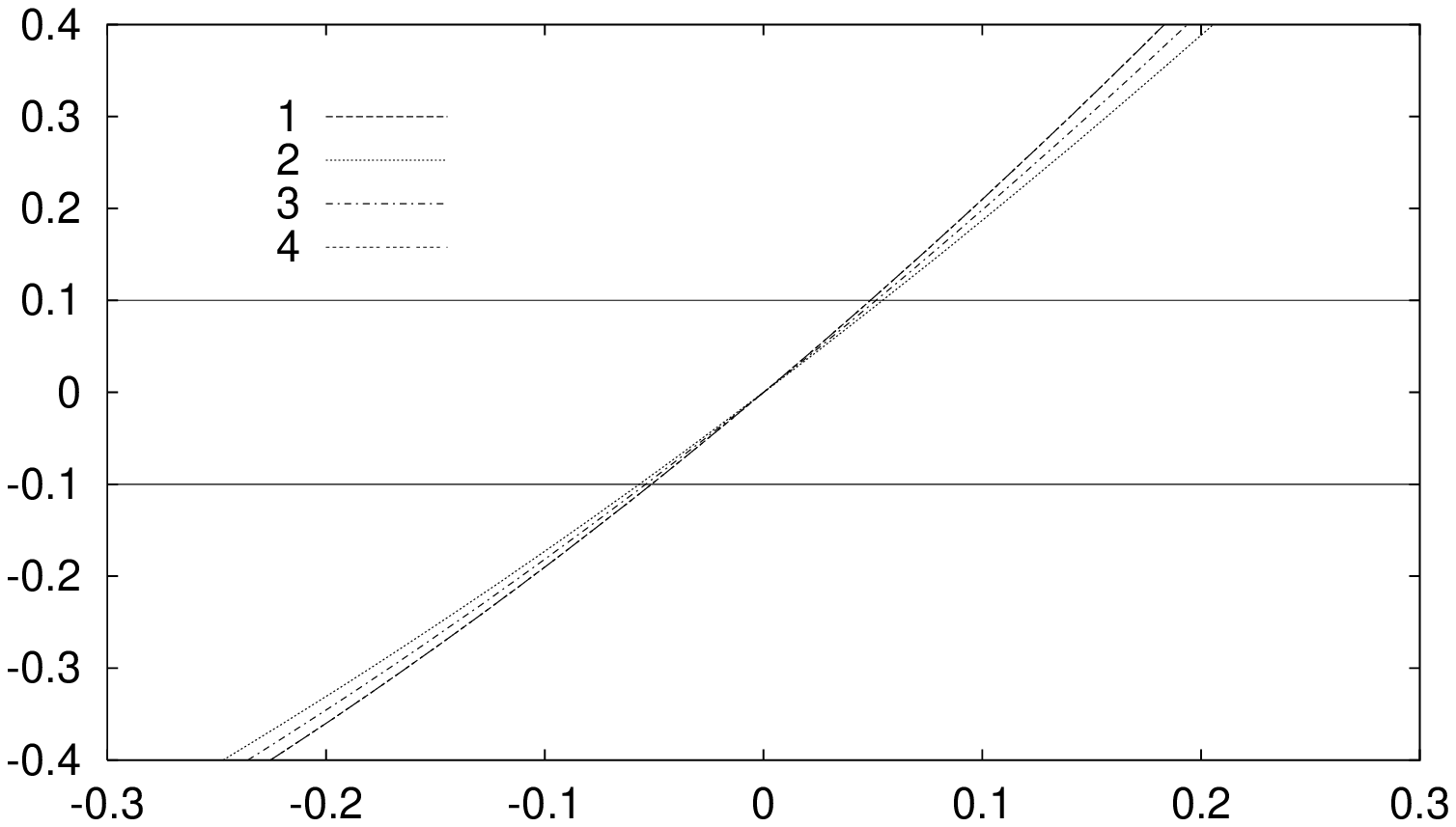}{\scalebox{1.2}{%
$\xi=(y^{+}-y_{p}^{+})/y_{p}^{+}$}}
{}}}
\put(25,150){\rotatebox{90}{\scalebox{1.05}{%
$(\upd \tau^+/\upd y^+)\boldsymbol{/}(\upd^2 U^+/\upd {y^+}^2)$}}}
\put(80,080){\scalebox{1.1}{1) $2\xi + \xi^{2}$}}
\put(80,060){\scalebox{1.1}{2) $(1+\xi)^{1.8}-1$}}
\put(80,040){\scalebox{1.1}{3) $(1+\xi)^{1.9}-1$}}
\put(80,020){\scalebox{1.1}{4) $(1+\xi)^{1.95}-1$}}
\put(190,080){\scalebox{1.05}{Universal logarithm law}}
\put(165,035){\scalebox{3.5}{$\left. \right\}$}}
\put(190,042){\scalebox{1.05}{Barenblatt profile}}
\end{picture}
\custcap{The ratio of the Reynolds shear stress gradient to the
viscous stress gradient in region II around $y_p$. The four curves
are obtained for different fits to the mean velocity data, as
explained at the bottom of the figure. The ratio for the Barenblatt
profile is given by $(1+\xi)^\alpha - 1$ where $\alpha=3/(2 \ln Re)$.}
\end{figure}
%
How large is this neighborhood? To answer this question, we plot in
Fig. 3 the ratio of the Reynolds shear stress gradient term to the viscous
term in the vicinity of $y_p^+$. Using experimental data for the
purpose would have generated much scatter, so our estimates are based
on various fits to the data: the classical logarithmic profile and
various power-law profiles recommended by~\cite{barenblatt93}. These
estimates suggest that the turbulent stress gradient term is of the
order of a tenth of the viscous stress gradient term within the
region $-0.1 \le \xi \le 0.1$, where $\xi = (y^+ - y_p^+)/y_p^+$. In
an order of magnitude sense, the region around $y_p^+$ where viscous
terms overwhelm turbulence terms has itself a width  $O(R_*^{1/2})$.

\section {The momentum equation}
\label{sec:analy}
Integrating Eq.~(\ref{eq:mombal}) and applying
the boundary condition at $y$ = 0 to
obtain the constant of integration, and that at $h$ to eliminate the
pressure gradient term, one obtains
\begin{equation}
\frac{\upd U^{+}}{\upd y^{+}} + \tau^+ =1-\frac{y^{+}}{R_{*}}.
\label{eq:meanmom}
\end{equation}
It appears natural that we should use Eq.~(\ref{eq:meanmom}) to
understand the mean
velocity distribution by approximating $\tau^+$ in the form of a
double expansion around $y_p^+$ and about the supremum value of unity
(attained in the infinite Reynolds number limit). We have accordingly
performed a local analysis
(local in $y^{+}$), in the limit
$R_{*}\rightarrow \infty$, of Eq.~(\ref{eq:meanmom}) in regions I and II
and III shown in the typical $\tau^+$
profile of Fig. 1. The regions are defined as domains of validity
of the asymptotic expansion of $dU^+/dy^+$ to the leading order in
the following limits:
\begin{eqnarray}
&R_{*}\rightarrow \infty,\:y^{+}=O(1)\qquad
&\mbox{(region I),} \nonumber \\
&R_{*}\rightarrow \infty,\:y^{+}=
O(y_{p}^{+}) \qquad &\mbox{(region II),} \nonumber \\
&R_{*}\rightarrow \infty,\:y^{+}=
O(R_{*}) \qquad &\mbox{(region III).} \nonumber
\end{eqnarray}
I is the classical viscous region, and III can be thought to be the
classical outer region.
Viscous terms are significant in both I and II, but the classical
buffer region, in which these terms are small in relative magnitude,
is interposed between the two regions. This makes the two regions
distinct. Here we shall focus on II nearly entirely.
The analysis of regions I and III in the same spirit yields
some significantly new results~\cite[]{sahay96b} which will
be summarized only as needed.

\subsection{Analysis of region II}
\label{subsec:analyII}
We can write the Taylor series expansion of
$\tau^+$ about $y_{p}^{+}$ as
$$
\tau^+(y^{+}) = \tau^+_{max} + \sum_{n=2}^{\infty}
f_{n}(R_{*})\xi^{n}
$$
where $\xi=(y^{+}-y_{p}^{+})/y_{p}^{+}$,
$n!f_{n}/{y_{p}^{+}}^n=[\upd^{n}\tau^+/\upd^n
y^{+}](y^{+}$=$y_{p}^{+})$, $f_{n}=o(R_{*}^\gamma)$
where $\gamma$ is any positive number, and $\tau^+_{max}$ is the
maximum value of $\tau^+$ for any given $R_*$. Clearly, $\tau^+_{max}
\le 1$, the equality holding only at infinite Reynolds number. For
any finite Reynolds number, we write $\tau^+_{max} = 1 - \Delta(R_*)$
(see Figs. 1 and 2), where $\Delta=o(1)$.  We may thus rewrite the above
equation as
\begin{equation}
\tau^+(y^{+}) = [1- \Delta(R_{*})] + \sum_{n=2}^{\infty}
f_{n}(R_{*})\xi^{n}.
\label{eq:uvtayII}
\end{equation}
Putting~(\ref{eq:uvtayII}) into~(\ref{eq:meanmom}) we get
\begin{equation}
\frac{\upd U^{+}}{\upd y^{+}}=
\Delta(R_{*}) - \frac{y_{p}^{+}}{R_{*}} - \frac{y_{p}^{+}}{R_{*}} \xi
+ \sum_{n=2}^{\infty}f_{n}(R_{*})\xi^{n}.
\label{eq:UtayIIa}
\end{equation}
If we take $\upd U^{+}/\upd y^{+}$ to be positive and monotonic
with respect to $y^{+}$ (which certainly seems to be case
empirically, although a theoretical proof is lacking),
we may argue that $\Delta=O(y_{p}^{+}/
R_{*})$: positivity implies $\Delta
R_{*}/y_{p}^{+}=o(R_{*}^{\gamma})$
for any positive $\gamma$ and monotonicity implies
$y_{p}^{+}/\Delta R_{*}=o(R_{*}^{\gamma})$.

The result that $\Delta=O(y_{p}^{+}/R_{*})$, when used in conjunction
with the estimate of $y_{p}^{+}$ (Eq. (4)), yields $\Delta =
O(R_{*}^{-1/2})$, say $a R_*^{-1/2}$, where $a$ is a constant. (This
latter relation is verified independently also by experiment.) Using
these estimates of the leading order of $y_{p}^{+}$ and $\Delta$ we
get,
from Eq.~(\ref{eq:UtayIIa}),
\begin{equation}
\frac{\upd U^{+}}{\upd y^{+}}=\frac{a}{R_{*}^{1/2}}-
\frac{\lambda}{R_{*}^{1/2}}-\frac{\lambda}{R_{*}^{1/2}} \xi +
o(R_{*}^{-1/2})[1+\xi] + \sum_{n=2}^{\infty}f_{n}(R_{*})\xi^n.
\label{eq:UtayIIb}
\end{equation}
Making use of the relation $y^{+}=y_{p}^{+}(1+\xi)=
\lambda R_{*}^{1/2} (1+\xi)$ we can rearrange~(\ref{eq:UtayIIb})
to obtain
\begin{equation}
\frac{\upd U^{+}}{\upd y^{+}}=\frac{(a-\lambda)\lambda}{y^{+}}+
\frac{(a-2\lambda)\lambda}{y^{+}}\xi+
o(R_{*}^{-1/2})[1+\xi]+\sum_{n=2}^{\infty}f_{n}(R_{*})\xi^{n}.
\label{eq:UtayIIc}
\end{equation}

\subsection{The logarithmic `law'}
An exact logarithmic profile would obtain if,
in a nonzero neighborhood of $\xi=0$, the first term
in Eq.~(\ref{eq:UtayIIc}) is dominant in the
limit $R_{*}\rightarrow \infty, y^{+}=O(y_{p}^{+})$.
The second term is of the order of the
first term and thus can be neglected only if it is identically
zero, which requires the exact equality of $a$ and $2\lambda$. This
assessment requires a numerical estimate of the constant $a$, which
we examine in Fig. 4.
%
\renewcommand{\plfig}[3]{
\begin{tabular}{c@{\hspace{1ex}}c}
\rotatebox[origin=c]{0}{#3} &
\raisebox{-0.5\height}{\makebox{\includegraphics[trim=10 0 10 0,%
width=0.92\textwidth,height=0.38\textheight]{#1}}}\\
 & #2
\end{tabular}}
\begin{figure}[t]
\centering
\begin{picture}(420,220)(45,-67)
\put(40,040){%
\mbox{\plfig{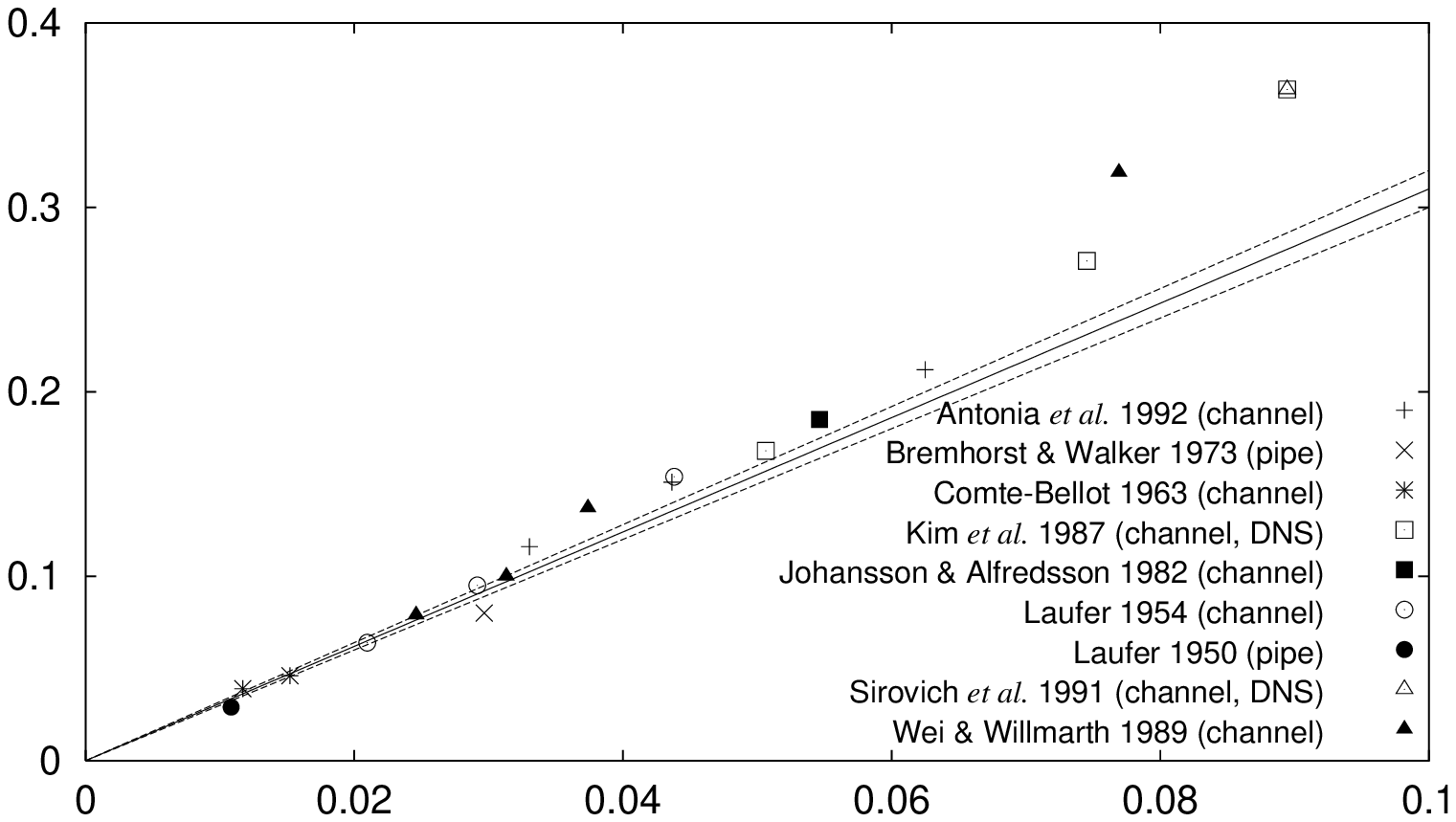}{\scalebox{1.15}{$R_{*}^{-1/2}$}}
{}}}
\put(43,000){\rotatebox{90}{\scalebox{1.15}
{$\Delta=1-\tau_{\mbox{max}}^{+}$}}}
\put(410,076){\vector(0,1){15}}
\put(410,076){\vector(-1,1){13}}
\put(410,076){\vector(1,1){15}}
\put(390,070){\scalebox{0.95}{$\Delta=a R_{*}^{-1/2}$}}
\put(383,060){\scalebox{0.95}{$a=3.1\pm0.1$}}
\renewcommand{\plfig}[3]{
\begin{tabular}{c@{\hspace{1ex}}c}
\rotatebox[origin=c]{0}{#3} &
\raisebox{-0.5\height}{\makebox{\includegraphics[trim=10 0 10 0,%
width=0.36\textwidth,height=0.17\textheight]{#1}}}\\
 & #2
\end{tabular}}
\put(90,090){%
\mbox{\plfig{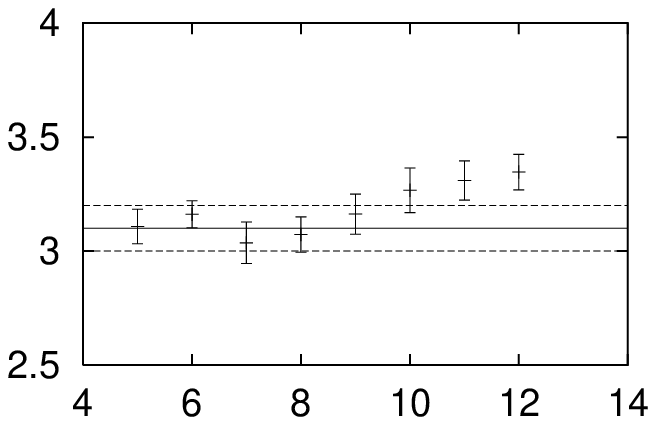}{}{}}}
\put(100,100){\scalebox{1.15}{$a$}}
\put(180,040){\scalebox{1.15}{$n$}}
\end{picture}
\custcap{Reynolds number variation of the leading order correction
to the peak value of $\tau^+$. The convergence of the estimate of $a$
was checked by varying the number of data-points ($n$ points
corresponding to the largest $R_{*}$) used for the least-square fit.
The variation of $a$ with $n$ is shown in the inset plot.
The error bars represent 68\% confidence intervals.}
\label{fig:uvyp}
\end{figure}
%

Here again it is difficult to determine
the coefficient accurately owing to finite-Reynolds-number
corrections. Ideally, one would consider data in intervals like
$[R_{*},\infty)$ for increasing $R_{*}$ and look for convergence
of the least-square estimates of the coefficient $a$.
The inset in Fig. 4
shows our attempt at implementing this algorithm within the
constraints
of finite maximum $R_{*}$ and small number of data points.
Least-square estimates of $a$ are obtained using $n$ data points at
the
largest available Reynolds numbers. We have taken into account only
the data
for $R_{*}\ge 500$. We do not consider smaller than five points
(i.e., we stop at $n=5$) since the fit for a smaller value of $n$
will lead to unacceptable statistical uncertainties. The resulting
estimate for $a$ is $3.1\pm0.1$ (see inset).

The uncertainty in the numerical estimates of $a$ and $\lambda$ allow
for the possibility that $a-2\lambda \equiv 0$, although the use of
their mean values makes the second term nonzero.
It is clear that empirical estimates will not settle
the issue of {\em exact} equality of the two numbers. An added
ignorance factor
is the lack of knowledge of the asymptotic
behavior of the functions $f_{n}(R_{*})$. For the logarithmic law to
exist,
the $f_{n}$ (derivatives of $\tau^+$ at $y_{p}$) must be $o(\Delta)$.

In spite of the ambiguity regarding the existence of an {\em exact}
logarithm law in region II, it is
true that there exists a nonzero neighborhood around $y_{p}^{+}$ in
which
a logarithmic variation is a {\em good approximation}
to the true velocity profile. In the extended limit
$R_{*}\rightarrow\infty, y^{+}=O(y_p^+), \xi\rightarrow 0$, the
leading term representation of~(\ref{eq:UtayIIc}) is
\begin{equation}
\frac{\upd U^{+}}{\upd y^{+}}=\frac{(a-\lambda)\lambda}{y^{+}},
\end{equation}
just as required for a logarithmic behavior. The log-law constant
$1/\kappa$ will then be $2.34\pm0.8$. Despite the large
uncertainty, the mean is very close to the traditionally accepted
value. A detailed analysis of region III of Fig.~\ref{fig:uvprof}~%
\cite[]{sahay96b} shows, however, that the outer edge of this
logarithmic region is $o(1)$.

\subsection{Local structure of the power-law profile}
\label{sec:barenb}
The expression for the profile is~\cite[]{barenblatt93,barenblatt96}
\begin{equation}
U^{+}=\beta {y^{+}}^{\alpha}, \qquad\qquad%
\beta=\frac{1}{\sqrt{3}}\ln Re + \frac{5}{2},\quad%
\alpha=\frac{3}{2\ln Re},
\label{eq:bareneq}
\end{equation}
where $Re$ is the Reynolds number based on the pipe diameter and the
average velocity across
the pipe cross-section. The factor $\sqrt 3$ in $\beta$ is an
aesthetic choice of the originator of the equation. The basis for
the choice of $\ln Re$ in $\alpha$ is that $\ln Re$ is insensitive to
the precise definition of $Re$ (see, e.g., \citealt{barenblatt95}).
The relationship between
$Re$ and $R_{*}$ is implicit in~(\ref{eq:bareneq}) and has been shown
to be~\cite[]{barenblatt93}
\begin{equation}
R_{*} = \frac{1}{2} \left[ \frac{e^{3/2 \alpha}2^{\alpha}\alpha(1+\alpha)%
(2+\alpha)}{\sqrt{3}+5\alpha} \right]^{1/(1+\alpha)}.
\label{eq:reyrel}
\end{equation}

Consider the limit process $R_{*}\rightarrow \infty,
y^{+}=O(y_{a}^{+})$
where $y_{a}^{+}$ is arbitrary. Let
$\zeta=(y^{+}-y_{a}^{+})/y_{a}^{+}$.
We can write Eq.~(\ref{eq:bareneq}) as
\begin{equation}
U^{+}=\beta\; \exp^{\alpha \ln y_{a}^{+}}\; \exp^{\alpha
\ln(1+\zeta)}.
\label{eq:bareneqII}
\end{equation}
Since $\zeta$ is $O(1)$ and $\alpha$ is $O((\ln
Re)^{-1})$
the second exponential can be expanded to yield, after some
rearrangement,
\begin{equation}
U^{+}=(\alpha \beta {y_{a}^{+}}^{\alpha}) \ln y^{+} +%
\beta {y_{a}^{+}}^{\alpha} (1-\ln {y_{a}^{+}}^{\alpha}) +%
O((\alpha \ln(1+\zeta))^{2}).
\label{eq:bareneqIII}
\end{equation}
Thus in a small neighborhood around $y_{a}^{+}$ the mean
velocity is like a logarithm. The asymptotic forms of the slope
($A\equiv \alpha \beta {y_{a}^{+}}^{\alpha}$) and the constant
($C\equiv \beta {y_{a}^{+}}^{\alpha} (1-\ln {y_{a}^{+}}^{\alpha})$)
will depend upon the order and magnitude of $y_{a}^{+}$.
If we take $y_{a}^{+}=a R_{*}^{\gamma}$ then it follows that
\begin{align}
A&=\exp^{3\gamma/2}\left[ \frac{\sqrt{3}}{2} +
O(\frac{\ln^2 R_{*}}{\ln R_{*}})\right] \\
C&=\exp^{3\gamma/2}\left[ \frac{1}{\sqrt{3}}%
\left( 1-\frac{3\gamma}{2} \right) \ln R_{*} + O(\ln^{2} R_{*}) \right]
\end{align}
where $\ln^{2} R_{*}=\ln(\ln R_{*})$.
Putting $\gamma=1$ the logarithmic variation predicted
for region III \cite[]{barenblatt96}
is obtained. The slope of this logarithmic law is indeed
${\sqrt{e}}$ times larger than the {\em universal} log-law.

It is of interest to examine the local structure of
Eq.~(\ref{eq:bareneq}) near $y_{p}^{+}$.
The local logarithmic
approximation~(\ref{eq:bareneqIII}) will hold with $y_{a}^{+}$
replaced by $y_{p}^{+}$. To calculate the asymptotic forms of the
slope and the constant (labeled $A_{p}$ and $C_{p}$ respectively)
we need an expression
for $y_{p}^{+}(R_{*})$ which is consistent with the power-law
expression for $U^{+}$, Eq.~(\ref{eq:bareneq}).
This can be done by substituting~(\ref{eq:bareneq}) in the mean
momentum equation~(\ref{eq:meanmom}) and solving for
$y^{+}=y_{p}^{+}$ at which
$\upd \tau^+/\upd y^{+}=0$. These manipulations yield
\begin{equation}
y_{p}^{+}=\biggl( (1-\alpha) \alpha \beta R_{*}
\biggr)^{1/(2-\alpha)}.
\label{eq:barenyp}
\end{equation}
In the limit $R_{*}\rightarrow \infty$, Eq.~(\ref{eq:barenyp}) can be
written
as
\begin{equation}
y_{p}^{+}=R_{*}^{1/2} \exp^{3/8} \left( \frac{\sqrt{3}}{2}
\right)^{1/2}%
\left[ 1+ \sum_{n=1}^{\infty} \sum_{m=0}^{n} P_{mn}%
\frac{(\ln^{2} R_{*})^{m}}{(\ln R_{*})^{n}} \right]
\label{eq:barenypII}
\end{equation}
where $P_{mn}$ are constants. The asymptotic forms of $A_{p}$ and
$C_{p}$ can easily be obtained.

The form of $U^{+}$ at $y_{p}^{+}$ corresponding to the
power-law~(\ref{eq:bareneq})
can be derived in a rational way from the mean momentum
equation, along the lines of the analysis in section 3.1.
We take into account (possible) corrections
to the leading order of $y_{p}^{+}$ and assume
\begin{equation}
y_{p}^{+}=\lambda R_{*}^{1/2}
\left[ 1+ \sum_{n=1}^{\infty} \sum_{m=0}^{n} \lambda_{mn}%
\frac{(\ln^{2} R_{*})^{m}}{(\ln R_{*})^{n}} \right]
\label{eq:ypgen}
\end{equation}
in accordance with~(\ref{eq:barenypII}). It is prudent to resolve
$\Delta$ to the
same degree as $y_{p}^{+}$, hence we take
\begin{equation}
\Delta=a R_{*}^{-1/2}
\left[ 1+ \sum_{n=1}^{\infty} \sum_{m=0}^{n} a_{mn}%
\frac{(\ln^{2} R_{*})^{m}}{(\ln R_{*})^{n}} \right].
\label{eq:delgen}
\end{equation}
The logarithmic `correction terms' to the leading order
dependence of $\Delta$ on $R_*$ are not unrealistic. For example,
returning to Fig.~\ref{fig:uvyp}, it is seen that
the $R_*^{-1/2}$ dependence is good only at very large
$R_*$. In Fig.~\ref{fig:uvypcr}, we show the difference $\Delta -
aR_*^{-1/2}$, which itself behaves to the leading order as
$\ln^2 R_{*}/(R_*^{1/2}\ln R_{*})$,
just as supposed in the expansion~(\ref{eq:delgen}).
%
%
\renewcommand{\plfig}[3]{
\begin{tabular}{c@{\hspace{1ex}}c}
\rotatebox[origin=c]{0}{#3} &
\raisebox{-0.5\height}{\makebox{\includegraphics[trim=10 0 20 0,%
width=0.92\textwidth,height=0.36\textheight]{#1}}}\\
 & #2
\end{tabular}}
\begin{figure}
\begin{picture}(420,210)(45,-60)
\put(40,050){%
\mbox{\plfig{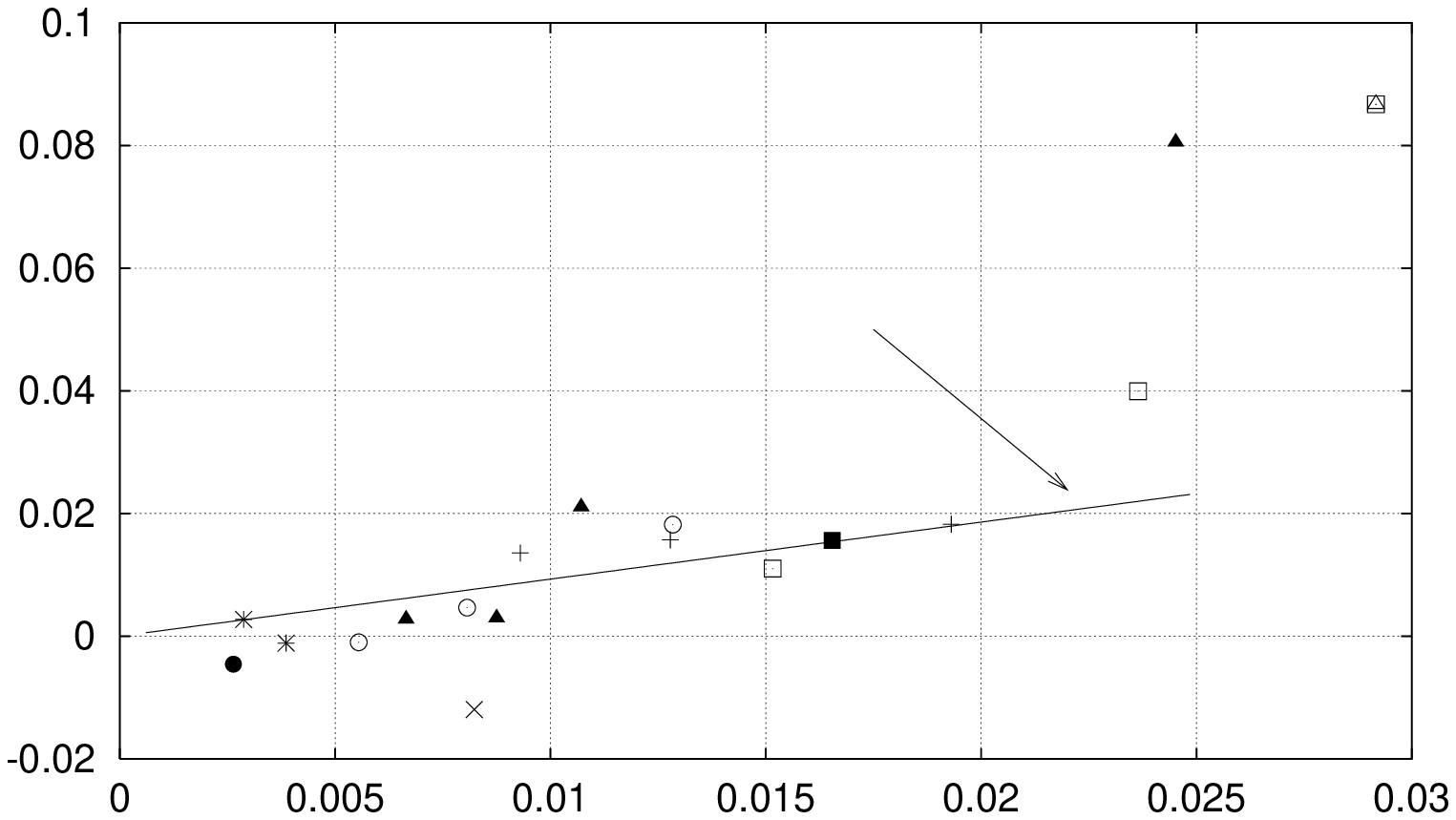}{}{}}}
\put(40,030){\rotatebox{90}{\scalebox{1.15}{$\Delta-3.1 R_{*}^{-1/2}$}}}
\put(210,-59){\scalebox{1.15}{$\ln^2 R_{*}/(R_{*}^{1/2}\ln R_{*})$}}
\put(240,80){\scalebox{1.1}{Slope=0.93}}
\end{picture}
\custcap{The difference between $\Delta$ and the leading order
behavior $3.1R_*^{-1/2}$, plotted against
$\ln^2 R_{*}/(R_*^{1/2} \ln R_*)$, in
order to determine the next-order term in Eq.~(\ref{eq:delgen}).
Using the best
fit to the data (except for those at the lowest three Reynolds
numbers) yields the relation $\Delta = 3.1R_*^{-1/2}
(1 + 0.3\ln^2 R_{*}/\ln R_{*})$. The symbols key is identical to
that of Fig. 4.}
\label{fig:uvypcr}
\end{figure}
%

Using Eqs.~(\ref{eq:ypgen}),~(\ref{eq:delgen}) and the relation
$y^{+}=y_{p}^{+}
(1+\zeta)$, we obtain an expression similar to~(\ref{eq:UtayIIc}) for
the
velocity derivate
\begin{align}
\frac{\upd U^{+}}{\upd y^{+}} &= \frac{a \lambda}{y^{+}}(1+\zeta)
\left[ 1+ \sum_{n=1}^{\infty} \sum_{m=0}^{n} a_{mn}^{\prime}%
\frac{(\ln^{2} R_{*})^{m}}{(\ln R_{*})^{n}} \right] \nonumber \\
&\quad -\frac{\lambda^{2}(1+\zeta)^{2}}{y^{+}}
\left[ 1+ \sum_{n=1}^{\infty} \sum_{m=0}^{n} \lambda_{mn}^{\prime}%
\frac{(\ln^{2} R_{*})^{m}}{(\ln R_{*})^{n}} \right]
+\sum_{n=2}^{\infty} f_{n} \zeta^{n}
\label{eq:UtayIIgen}
\end{align}
where $a_{mn}^{\prime}$ and $\lambda_{mn}^{\prime}$ are constants
related to $a_{mn}$ and $\lambda_{mn}$.
In the extended limit $R_{*}\rightarrow \infty, y^{+}=O(y_{p}^{+}),
\zeta \rightarrow 0$ the leading term
of~(\ref{eq:UtayIIgen}) is of the same form as that of
$\upd U^{+}/\upd y^{+}$ for the Barenblatt profile.

In summary, it appears that the Barenblatt--Chorin profile is consistent with
Eq.~(\ref{eq:meanmom}) when the double series expansions~(\ref{eq:ypgen}) and
(\ref{eq:delgen}) are used for $y^+_p$ and $\Delta$.

\section{Conclusions}

The primary qualitative point of this work is that there exists a
mechanism for viscous effects to spread to the classical overlap
region. The existence of such a mechanism will in principle prevent
classical matching, so that the leading order `inner' and `outer'
expansions are technically insufficient to construct a uniformly
valid approximation of the mean velocity profile.

What, then, can be said about the classical log-law? First, it must
be noted that the velocity change across the `mesolayer' is not of
the order unity, unlike in most boundary layer problems, which may
make its importance potentially less significant. Second, just as is
done here for region II, we have analyzed regions I and III in some
detail, making use of the well-known forms of the Reynolds shear stress
distributions in those regions. (For example, in region III, one has
a linear distribution of the Reynolds shear stress, with some
well-understood finite Reynolds number corrections.) It is then
possible to patch together the results from all three regions.
A summary of this work is as follows. In region I, beyond about $y^+
\sim 30$, $dU^+/dy^+$ lies close to that implied by the classical
logarithmic law. In region II, we have already seen that the log-law
can be a good approximation (and even exact in a small neighborhood
of $y_p^+$). In region III, $dU^+/dy^+$ lies close to the classical
value for $y^+$ smaller than $0.15R_*$. It is therefore reasonable
that one can smoothly match the three regions and obtain a
logarithmic variation of velocity between $y^+ \sim 30$ and  $y^+
\sim 0.15R_*$, regardless of whether it is exact.

What can be said of power-laws, in particular the Barenblatt profile?
In our present view, a self-consistent way of understanding the
origin of this profile is to use Eq.~(\ref{eq:meanmom}) and the appropriate
expansions for $\Delta$ and $y_p$ in B1 (and the appropriate version
of $y_a$ in region B2). We believe that the rich structure of
power-law profiles makes them
fit the data well. However, a close analysis of region III
\cite[]{sahay96b} reveals that asymptotically the velocity
profile cannot be exactly logarithmic in B2. The most
significant point in favor of power laws is that they account, in
some fashion, for viscous effects to pervade in the classical overlap
region.

The importance of viscous effects in a region traditionally thought
to be inviscid has other obvious analogies. The principal analogy is
the Kolmogorov spectral cascade. Just as viscous effects are centered
around the peak position of the Reynolds shear stress in the present
problem, one may imagine the viscous effects in Kolmogorov turbulence
may be centered around the position of maximum energy transfer in
wave number space. This provides a natural mechanism for viscous
effects to encroach the inertial region of the spectral space.
Details are currently being worked out.

We have so far not speculated on the physical origin of the
viscous effects in the neighborhood of $y_p$. Two possible scenarios
present themselves. It may well be that the bursting of wall-layer
streaks (which undoubtedly extend beyond the sublayer) carry viscous
effects with them as they penetrate up to $y^+ = O(R_*^{1/2})$.
Alternatively, the viscous cores of the coherent vortices in the
`critical layer' region are such that their effects do not vanish at
any Reynolds number (Barenblatt 1993). Whatever the mechanism,
it appears (Sreenivasan 1987) that the `critical layer' or the
mesolayer plays an important role in the dynamics of wall flows that
cannot be subsumed in the classical picture.

\section*{Acknowledgements}
The work was supported by a grant from AFOSR.

\nocite{antonia92,bremhorst73,comte63,johansson82,kim87}
\nocite{lauf50,lauf54,sirov91,wei89}
\tenrm
\bibliographystyle{../../bibtex/contrib/natbib/natbibpers}
\bibliography{../channel}
\newpage
\end{document}